\documentclass[11pt,reqno]{amsart}
\usepackage{amssymb,mathrsfs,color}
\usepackage{pinlabel}

\usepackage{amsmath, amsfonts, amsthm, verbatim}
\usepackage{epstopdf, mathtools}
\mathtoolsset{showonlyrefs}
\usepackage{color}
\usepackage{esint}
\usepackage{stmaryrd}

\usepackage{empheq}
\usepackage[shortlabels]{enumitem}

\newcommand{\rar}{\rightarrow}

\newcommand{\cl}{\mathcal}

\newcommand{\bs}[1]{\boldsymbol{#1}}

\newcommand{\spn}{\operatorname{span}}

\renewcommand{\ker}{\operatorname{ker}}

\definecolor{deepgreen}{cmyk}{1,0,1,0.5}

\newcommand{\al}{\alpha}

\newcommand{\om}{\omega}
\newcommand{\la}{\lambda}

\newcommand{\p}{\partial}

\makeatletter

\newcommand{\Rmnum}[1]{\expandafter\@slowromancap\romannumeral #1@}
\makeatother

\newcommand{\ang}[1]{\left\langle{#1}\right\rangle}

\newcommand{\Del}[1]{}

\numberwithin{equation}{section}

\newtheorem{thm}{Theorem}[section]

\newtheorem{lem}[thm]{Lemma}
\newtheorem{prop}[thm]{Proposition}

\theoremstyle{remark}
\newtheorem{rem}[thm]{Remark}

\renewcommand\Re{\mathrm{Re}\,}

\usepackage{tensor}

\usepackage{graphicx}
\usepackage{xcolor} 
\usepackage{tensor}
\usepackage{slashed}
\usepackage{cite}
\definecolor{green}{rgb}{0,0.8,0} 

\usepackage[bottom]{footmisc}

\renewcommand{\Re}{\mathrm{Re}}

\newcommand{\eps}{\epsilon}

\newcommand{\bbC}{\mathbb C}

\newcommand{\bbE}{\mathbb E}

\newcommand{\bbN}{\mathbb N}

\newcommand{\bbR}{\mathbb R}

\newcommand{\mucir}{\mu^\circ}

\newcommand{\spa}{\mbox{span}}

\newcommand{\nucir}{\nu^\circ}

\begin{document}

\title[Evolving natural curvature]{On evolving natural curvature for an inextensible, unshearable, viscoelastic rod}

\author{K. R. Rajagopal and C. Rodriguez}

\dedicatory{Dedicated to the memory of Prof. J. L. Ericksen.}

\begin{abstract}
We formulate and consider the problem of an inextensible, unshearable, viscoelastic rod, with evolving natural configuration, moving on a plane. We prove that the dynamic equations describing quasistatic motion of an Eulerian strut, an infinite dimensional dynamical system, are globally well-posed. For every value of the terminal thrust, these equations contain a smooth embedded curve of static solutions (equilibrium points). We characterize the spectrum of the linearized equations about an arbitrary equilibrium point, and using this information and a convergence result for dynamical systems due to Brunovsk\'y and Pol\'acik, we prove that every solution to the quasistatic equations of motion converges to an equilibrium point as time goes to infinity. 
	
\end{abstract}

\maketitle

\section{Introduction}

In an influential paper concerning the equilibrium of rods, Ericksen \cite{EricksenBars75} considered the equilibrium states of a straight, extensible Green elastic rod, whose constitutive relation for the stored energy is a non-convex function of the deformation gradient.\footnote{The more geometric problem of bending of an inextensible, unshearable rod  whose constitutive relation for the stored energy is a non-convex function of the rod's curvature was later treated by Fosdick and James \cite{FosdickJames81}.} This authoritative paper has motivated numerous studies which purport to study phase transition in solids but are really concerned with equilibrium states composed of co-existent parts with distinct deformation gradients. The problem studied by Ericksen concerns the equilibrium of a rod with juxtaposed parts with distinct constant deformation gradients, a purely static consideration and a purely kinematic consideration.\footnote{A lengthy discussion as to why these studies do not concern the actual process of phase transition but are merely concerned with a static deformed body comprised of elastic bodies with two distinct microstructures juxtaposed to each other in equilibrium, that is, they are concerned with the arrangement of a mixture of elastic materials with no phase change taking place, can be found in Rajagopal and Srinivasa \cite{RajSriZAMP04a}.} 

The one-sentence abstract to his paper states “For elastic bars, we discuss some material instabilities”, the study that is carried out is purely within the purview of elastostatics. Ericksen recognizes that in order to discuss “instabilities” one needs to study the problem within a dynamical context, as he states “For our purposes, it is essential to employ a stability criterion, and we employ the traditional energy criterion. As is well known, it is not completely reliable, but it is hard to improve upon it, without specifying constitutive equations which are to apply out of equilibrium.” 

A little later in the introduction he makes the comment “In one sense, ours is an elementary study of phase transformations that are shear induced, i.e., induced by stresses which are not hydrostatic pressures”, and follows this up subsequently with the remark “In the analog of van der Waals’ fluid, we would be in a range where a fixed volume contains a mixture of liquid and vapor phases, when the equation of state exhibits a similar oscillation” . This remark of Ericksen’s concerning the equilibrium of distinct phases has been picked up by numerous researchers who have generalized this study to what they refer to as the problem of phase transition in elastic solids, purely within an elastostatic context wherein the elastic body is described by a non-convex stored energy function. This is an unfortunate circumstance for a variety of reasons. First, just from the viewpoint of English, it is incorrect. The term “transition” refers to a dynamic situation. As one gathers from the Oxford English Dictionary \cite{OxfordDictionary2000}, the word “transition” means “A passing or passage from one condition, action, or (rarely) place, to another: change”. Thus, if one is to study phase transition, it has to be within a dynamical context. Second, “phase transitions” are entropy producing. Without knowing the constitutive relation for the rate of entropy production, and merely knowing how the material stores energy, we cannot describe the phenomenon of “phase transition”. Third, even when there is no “phase transition”, such energy-based studies do not necessarily work and Ericksen was clearly aware of it and mentions it in the comment quoted above. Fourth, just because the governing equations are similar in totally distinct physical problems, one cannot conclude the physical quantities being described share identical attributes. Such flawed thinking can lead to disastrous consequences in physics. There are many more reasons why one ought not to interpret Ericksen’s study as being relevant to “phase transformations”, but we shall not get into a discussion of them here. Ericksen’s study reveals that there are equilibrium states of a purely elastic rod, wherein the deformation gradient could be discontinuous, no more and no less. The lack of relevance to phase transition notwithstanding, Ericksen’s study was thought provoking and opened up a fertile area for further study in nonlinear elasticity.

There are two important attributes to an elastic body, it is incapable of dissipating energy, that is, converting work into energy in its thermal form (heat), and it has a unique “natural configuration”.  The notion of a natural configuration seems to have been discussed first by Eckart \cite{EckartThermo48}. One could view the natural configuration  of a body to be the configuration in which it would exist on the removal of all external stimuli.\footnote{The notion of a natural configuration is a local notion. When the external stimuli on a body that has been inhomogeneously deformed, are removed, the body might not attain a configuration which coheres properly. This is precisely the situation in an inelastic solid that has yielded. We can either consider the unloaded configuration in a non-Euclidean space, as considered by Eckart, or merely consider small locally homogeneously deformed neighborhoods that are unloaded in an Euclidean space as the local “natural configuration”.} From a purely mechanical perspective, it would be the configuration that the body would take if all the external loads were removed (see Rajagopal \cite{RajagopalReport695} for a discussion of the notion of natural configuration). 

Unlike elastic bodies that are characterized by a single natural configuration, bodies such as viscoelastic and inelastic bodies are characterized by multiple natural configurations, and the body’s natural configuration can evolve when the body is undergoing a thermodynamic process. The manner in which the natural configuration evolves is determined by the way in which entropy is produced by the body.  In bodies capable of producing entropy, entropy production could be due to various causes, due to dissipation (work being converted to thermal energy), due to conduction, due to mixing, due to phase transition, due to growth, etc. If we restrict our attention to purely mechanical processes, then the only source of entropy production is that due to dissipation. In the case of an elastic body, within a purely mechanical context, no entropy is produced, and hence its natural configuration does not evolve. However, in the case of a viscoelastic body, even within a purely mechanical context, the natural configuration can evolve, as the body is capable of producing entropy. The study of phase transition would have to take this evolution of the natural configuration of the body into consideration.

Based on these considerations of Ericksen's work, we study the problem of an inextensible, unshearable, viscoelastic rod with evolving natural configuration, moving on a plane, with one end fixed and the other end free. Because of the symmetry of the problem, the rod's configuration, a planar curve $s \mapsto \bs r(s,t)$, is determined uniquely\footnote{Uniqueness holds up to a rigid motion of the plane containing the center line of the rod.} by one scalar function; the rod's curvature $\mu(\cdot,t)$ at time $t$. At each time $t$, we posit a second configuration ${\bs r}^\circ(\cdot,t)$, referred to as the \emph{natural configuration}, with curvature $\mu^\circ(t)$. We refer to the curvature $\mu^\circ(t)$ determining the rod's natural configuration as its \emph{natural curvature}. Our terminology is based on the fact that in our model the contact couple vanishes throughout the rod, at time $t$, precisely when $\mu(\cdot,t) = \mu^\circ(t)$ and $\mu_t(\cdot,t) = 0$ (see \eqref{eq:natcondition}). 
 While it would be tempting to think of each natural configuration as characterizing a different ``phase" of the rod, it would not be sensible to do so as there are not really any different phases of the material present. 

Our viscoelastic rod is characterized by two scalar functions, a stored energy and a rate of total entropy production. The stored energy
\begin{align}
	\frac{(EI)}{2}(\mu(s,t) - \mu^\circ(t))^2 +
	\kappa(\mu^\circ(t))
\end{align}
 is the sum of two contributions. The first contribution is stored energy due to deformation. The second contribution is stored energy due to the rod's natural curvature $\mu^\circ(t)$ being different from other fixed preferred natural curvatures. For example, it seems plausible to assume that the preferred natural curvature of a virgin straight rod is zero. However, this object may be bent (with enough force) changing its current natural curvature to another value. With respect to entropy production and unlike most studies in continuum thermodynamics which appeal to the Clausius-Duhem inequality locally, we only require the condition that the \emph{total} entropy of the rod is non-increasing in time,
 \begin{align}
 	\frac{d}{dt} \int_0^L \eta(s,t) ds \geq 0,
 \end{align}
where $\eta(\cdot,t)$ is the entropy per unit reference length at time $t$.  In the purely mechanical setting that we consider, the rate of total entropy production has two contributions (see \eqref{eq:totalentropy}). The first contribution is due to the viscoelastic nature of the rod. The second contribution to the total entropy production is due to a difference in the averaged curvature of the rod's current configuration and the curvature of the rod's current natural configuration. The form of this term we posit determines the evolution equation for the natural curvature (see \eqref{eq:mueq}).  In particular, this second contribution to the total entropy production is positive if and only if the natural curvature is changing. Moreover, a nontrivial threshold of energy is required to produce a change in the rod's natural curvature.

In this work we consider the quasistatic motion of a viscoelastic rod, with evolving natural curvature, with one fixed end and with one free end subjected to a terminal thrust (an Eulerian strut). The structure and main results of our study are as follows. 

In Section 2 we first review the general equations of motion for an inextensible, unshearable special Cosserat rod moving on a plane. Next we develop the constitutive relations and evolution equation for the natural curvature $\mu^\circ(t)$ based on thermodynamic considerations. We emphasize here that we require that the total entropy of the rod, $\int_0^L \eta(s,t) ds$, rather than the pointwise value of the entropy $\eta(s,t)$, is increasing in time (see \eqref{eq:dissin}). A nice feature of this weaker requirement is that we can make the simplifying geometric and logically consistent assumption that the natural curvature $\mu^\circ$ is independent of the variable $s$ for each $t$.\footnote{Equivalently, we assume that the possible natural configurations of the rod are segments of circles of varying radii.} If instead we require the entropy to be pointwise increasing in time, then we could not make this assumption. Of course, many rods can and do have a non-uniform natural curvature evolving in time, and we intend to investigate the more general model in future work.  

In Section 3, we specialize our study to the equations describing quasistatic motion of an Eulerian strut (see \eqref{eq:muequation} and \eqref{eq:mucircequation}).  These equations can be written as a differential equation on $L^2(0,1) \times \bbR$ 
\begin{align}
	\p_t (\mu(s,t), \mu^\circ(t)) = F(\mu(\cdot,t), \mu^\circ(t))(s)
\end{align}
for the pair $(\mu(\cdot,t), \mu^\circ(t))$, where $F(\mu, \mu^\circ)$ is given by \eqref{eq:Fdef}.  Our thermodynamic considerations from Section 2 imply the existence of a Liapunov function for this infinite dimensional dynamical system (see \eqref{eq:liapunov}), and by the standard theory of differential equations in Banach spaces, we conclude global well-posedness for the quasistatic equations of motion (see Theorem \ref{p:wellposedness}). We then consider equilibrium points, $(\nu, \nu^\circ) \in L^2(0,1) \times \bbR$ satisfying $F(\nu, \nu^\circ) = (0,0)$. This subset of $L^2(0,1) \times \bbR$ always contains a smooth embedded curve through the trivial equilibrium point $(0,0)$.\footnote{The equilibrium point $(0,0)$ corresponds to the straight configuration of the rod.} We characterize the spectrum $\sigma(DF(\nu,\nu^\circ))$ of the linearization about a given equilibrium point $(\nu, \nu^\circ)$. In particular, $\sigma(DF(\nu,\nu^\circ)) = \sigma^u \cup \{0\} \cup \sigma^s$ where $\sigma^u$ is a finite subset of $(0,\infty)$, $\sigma^s$ is a closed subset of $(-\infty,0)$, and $0$ is a simple eigenvalue (see Proposition \ref{p:spectrum}).  In the generic case that $|D| < \Theta(\nucir)$, $\ker DF(\nu, \nu^\circ)$ is tangent to a smooth embedded curve through $(\nu,\nu^\circ)$ in $L^2(0,1) \times \bbR$ consisting entirely of equilibrium points (see Proposition \ref{p:curve}).  

Finally, in Section 4 we prove that an arbitrary solution to the quasistatic equations of motion converges to an equilibrium point, as $t \rar \infty$, in $L^2(0,1) \times \bbR$. We first show that the trajectory of a solution is precompact in $L^2(0,1) \times \bbR$ (see Lemma \ref{l:compact}). Next, we show that the set of limit points of the trajectory consists entirely of equilibrium points (see Lemma \ref{l:om}). Finally, using the spectral information from Section 3 (in particular that the center space is one dimensional) and a convergence result for dynamical systems due to Brunovsk\'y and Pol\'acik \cite{BPolacik97}, we show that the set of limit points of the trajectory is a singleton and conclude (see Theorem \ref{t:bp97}).

\section{Equations of Motion}

In this section we formulate the theory for an inextensible, unshearable rod with evolving (uniform) natural curvature, moving on a plane. For a complete introduction to rods in general, we refer the reader to the authoritative treatise by Antman \cite{AntmanBook}. 

\subsection{Preliminaries}
 
Let $\bbE^3$ be Euclidean space, and let $\{ \bs i, \bs j, \bs k \}$ be a fixed right-handed orthonormal basis for the vector space $\bbR^3$. We identify $\bbE^2 = \{ (x_1,x_2,0) \in \bbE^3 \mid x_1,x_2 \in \bbR \}$.  Let $[0,L]$ be the reference interval parameterizing the particles or material points of the rod. The planar \emph{configuration} of a uniform, inextensible, unshearable special Cosserat rod at time $t$ is given by a curve 
\begin{align}
	[0,L] \ni s \mapsto \bs r(s,t) \in \bbE^2. 
\end{align} 
with 
\begin{align}
	\bs r_s(s,t) &= \cos \theta(s,t) \bs i + \sin \theta(s,t) \bs j. 
\end{align}
The curve $\bs r(\cdot, t)$ is the center line of the rod and the vector  $$\bs b(s,t) = - \sin \theta(s,t) \bs i + \cos \theta(s,t) \bs j$$ is the director at $(s,t)$, a unit vector normal to $\bs r(\cdot,t)$ at $\bs r(s,t)$.  We set 
$$\bs a(s,t) = \bs r_s(s,t) = \cos \theta(s,t) \bs i + \sin \theta(s,t) \bs j,$$
so $\{ \bs a(s,t), \bs b(s,t)\}$ is a right-handed orthonormal basis for $\bbR^2$, for each $(s,t)$. We note that 
\begin{align*}
	\mu(s,t) = \theta_s(s,t)
\end{align*}
is the curvature of the curve $\bs r(\cdot,t)$ at $s$ and is the sole measure of strain for an inextensible, unshearable special Cosserat rod. 

At each time $t$ during the motion of the rod, we specify a scalar $\mu^\circ(t)$.  The specification of $\mu^\circ(t)$ determines a configuration $\bs r^\circ(\cdot,t)$ via $\bs r^\circ(s,t) = (s,0,0)$ if $\mu^\circ(t) = 0$ and 
\begin{align}
\bs r^\circ(s,t) = \Bigl (\frac{\sin (\mu^\circ(t) s)}{\mu^\circ(t)},
\frac{1- \cos (\mu(t) s)}{\mu^\circ(t)}, 0 \Bigr ), \quad s \in [0,L],
\end{align} 
if $\mu^\circ(t) \neq 0$. 
This configuration has curvature $\mucir(t)$ and is
unique up to a rigid displacement in $\bbE^2$. We refer to $\mucir(t)$ as the \emph{natural curvature} at time $t$ and $\bs r^{\circ}(\cdot,t)$ as the associated \emph{natural configuration} at time $t$. Why we refer to $\mucir(t)$ as the natural curvature is discussed in the next subsection (see \eqref{eq:natcondition}). In the classical theory of rods a single natural configuration is fixed for all time, $\mu^\circ(t) = \mu^\circ(0)$ for all $t$. In general micro-structural changes may result in $\mu^\circ(t) \neq \mu^\circ(0)$ at some $t > 0$ so that the natural configuration evolves in time. 

\subsection{Balance laws and entropy production}

Let $[a,b] \subseteq [0,L]$. We denote the contact force by $\bs n(s,t) \in \spa \{\bs i, \bs j\}$ so that at each time $t$, the resultant force on the material segment $[a,b]$ by $[0,a) \cup (b,L]$ is given by $$\bs n(b,t) - \bs n(a,t).$$ We denote the contact couple by $\bs m(s,t) \in \spa \{\bs k\}$ so that at each time $t$, the resultant contact couple about $\bs o =  (0,0,0)$ on the material segment $[a,b]$ by $[0,a) \cup (b,L]$ is given by 
$$\bs m(b,t) + (\bs r(b,t) - \bs o) \times \bs n(b,t) - \bs m(a,t) - (\bs r(a,t)-\bs o) \times \bs n(a,t).$$
If there are no external body forces or body couples, then the classical equations expressing balance of linear momentum and angular momentum are given by: for $(s,t) \in [0,L] \times (0,\infty)$
\begin{align}
	(\rho A) \bs r_{tt}(s,t) &= \bs n_s(s,t), \label{eq:lm} \\
	(\rho J) \bs b(s,t) \times \bs b_{tt}(s,t) &= \bs m_s(s,t) + \bs r_s(s,t) \times \bs n(s,t), \label{eq:am}
\end{align} 
where $(\rho A)$ and $(\rho J)$ are the constant mass and first moment of mass per unit reference length. Expressing $\bs n = N \bs a + H \bs b$, $\bs m = M \bs k$, we see that \eqref{eq:lm} and \eqref{eq:am} are equivalent to 
\begin{align}
	(\rho A) \bs r_{tt} \cdot \bs a &= N_s, \label{eq:Neq} \\
		(\rho A) \bs r_{tt} \cdot \bs b &= H_s, \label{eq:Heq} \\
			(\rho J) \theta_{tt} &= M_s + H \label{eq:Meq}. 
\end{align}
The functions $N$ and $H$ act as Lagrange multipliers enforcing the constraint $\bs r_s = \bs a$ and are unknowns. For $M$ we assume an Euler-Bernoulli relation: 
\begin{align}
	M(s,t) = (EI)[(\mu(s,t) - \mu^\circ(t)) + \nu \mu_t(s,t)], \label{eq:Mrel}
\end{align}
where $(EI) > 0$ and $\nu > 0$ are constants. In particular, 
\begin{align}
	M(\cdot,t) = 0 \iff \mu(\cdot,t) = \mu^\circ(t), \quad \p_t \mu(\cdot,t) = 0. \label{eq:natcondition}
\end{align}
It is \eqref{eq:natcondition} that encapsulated why we refer to $\mu^\circ(t)$ as the natural curvature at time $t$.
All that remains to close the system of equations is to posit an evolution equation for the curvature of the natural configuration, $\mu^\circ(s,t)$. This will be obtained via the following thermodynamic considerations. 

We assume that the rod is held at a fixed absolute temperature $\tau$ and is nonconducting, and the internal energy per unit reference length $e(s,t)$ is given by 
\begin{align}
	e(s,t) = \tau \eta(s,t) + \frac{(EI)}{2}(\mu(s,t) - \mu^\circ(t))^2 +
	\kappa(\mu^\circ(t))
	\label{eq:eeq}
\end{align} 
where $\eta(s,t)$ is the entropy per unit reference length, and $\kappa$ satisfies: 
\begin{itemize}
	\item $\kappa$ is twice continuously differentiable on $\bbR$, non-negative, and even, 
	\item $\kappa(\mu^\circ) \rar \infty$ as $|\mu^\circ| \rar \infty$,
		\item $\kappa(0) = \kappa'(0) = 0$. 
\end{itemize}
The final condition on $\kappa$ can be interpreted as the straight configuration being a preferred natural configuration of the rod. 
We remark that if in addition, 
\begin{itemize}
	\item there exists $\mu_1^\circ > 0$ such that $\kappa(\mu^\circ) = 0$ if and only if $\mu^\circ \in \{0, \pm \mu_1^\circ\}$, 
	\item $\kappa'(\pm \mu_1^\circ) = 0$, 
\end{itemize}
then we can formally interpret the configuration with $\mu^\circ \in [0,\mu^\circ_1]$ as a juxtaposition at the microscopic level of the two ``phases" determined by $\mu^\circ = 0$ and $\mu^\circ = 1$ as follows.\footnote{As discussed in the introduction, the rod with natural curvatures $0$ and $\mu_1^\circ$ should not really be considered as two different ``phases" since there are not really any different phases of the material present.}  For $\alpha \in [0,1]$, we define the configuration $\bs r_\alpha^\circ$ of the intermediate state with volume fraction $\alpha$ of the product phase via specifying its curvature:  
\begin{align}
	\mu^\circ_\alpha = (1-\alpha)0 + \alpha \mu_1^\circ = \alpha \mu_1^\circ.  
\end{align}
In particular, the configuration with $\mu^\circ \in [0, \mu^\circ_1]$ corresponds to an intermediate state with volume fraction $\alpha = \mu^\circ/\mu^{\circ}_1$. 
The strain of the configuration $\bs r(\cdot, t)$ relative to $\bs r_\alpha^\circ$ at $(s,t)$ is defined to be 
\begin{align*}
	\mu_\alpha(s,t) := \mu(s,t) - \mu^\circ_\al. 
\end{align*}
Then for $\mu^\circ(t) \in [0, \mu^\circ_1]$ with $\alpha(t) = \frac{\mu^\circ(t)}{\mu_1^\circ} \in [0,1]$ we can write the internal energy as 
\begin{align}
	e(s,t) = \tau \eta(s,t) + W(\mu_{\al(t)}(s,t)) + J(\alpha(t))
\end{align}
where $W(\mu_\al) := \frac{(EI)}{2}\mu_\al^2$ and $J(\alpha) := \kappa(\alpha \mu^\circ_1)$. This is the form of the internal energy used by Rajagopal and Srinivasa (see e.g. \cite{RAJSRI19951}, \cite{RAJSRI19952}, \cite{RAJSRIZAMP99}, \cite{RajSriZAMP04a}, \cite{RajSriZAMP04b})), and one can then obtain equations of motion in terms of $\alpha$ instead of $\mu^\circ$ (for $\mu^\circ \in [0, \mu_1^\circ]$). In this case, one should interpret the term $\kappa$ as accounting for energy due to the rod being a mixture of the two ``phases" with mismatching curvatures. 

The classical equation expressing balance of energy reduces to 
\begin{align}
	e_t(s,t) = M(s,t)\mu_t(s,t). \label{eq:enbal}
\end{align}
We assume that the total entropy is non-decreasing in time:  
\begin{align}
	\frac{d}{dt} \int_0^L \eta(s,t) ds = \int_0^L \eta_t(s,t) ds \geq 0. \label{eq:dissin}
\end{align}
We note that \eqref{eq:dissin} is weaker than the form most often assumed: $\eta_t(s,t) \geq 0$ for all $(s,t)$. Also, in the approach adopted by Green and Naghdi \cite{GreenNaghdi77a}, the entropy inequality is expressed as an equality, a constitutive relation is assumed for the entropy production, and the thermodynamic process is determined by maximizing the entropy production (see Rajagopal and Srinivasa \cite{RAJAGOPALSRI2000}). 
Then \eqref{eq:Mrel}, \eqref{eq:eeq} and \eqref{eq:enbal} imply that  
\begin{align}
	\int_0^L \eta_t(s,t) ds &= \int_0^L (EI)\nu (\mu_t(s,t))^2 ds + D(t) \mu_t^\circ(t), \label{eq:diss} \\
	D(t) &= \int_0^L (EI)(\mu(s,t) - \mu^\circ(t)) ds - L\kappa'(\mu^\circ(t)).  
\end{align}
We specify the form of the rate of total entropy production: 
\begin{gather}
	\int_0^L \eta_t(s,t)ds = \int_0^L (EI) \nu (\mu_t(s,t))^2 ds \\ + |D(t)| [f(D(t)-\Theta(\mu^\circ(t))) + f(-\Theta(\mu^\circ(t)) - D(t))], \label{eq:totalentropy}
\end{gather} 
where
\begin{itemize}
	\item $f$ is continuously differentiable on $\bbR$ and non-negative, 
	\item if $x \leq 0$ then $f(x) = 0$, 
\end{itemize}
and 
\begin{itemize}
	\item $\Theta$ is continuously differentiable on $\bbR$, positive and even, 
	\item $\Theta$ is bounded from below, 
\begin{align}
	\inf_{\mu^\circ \in \bbR} \Theta(\mu^\circ) > 0. \label{eq:Alower}
\end{align}
\end{itemize}

Then \eqref{eq:dissin} is automatically ensured and \eqref{eq:diss} is equivalent to 
\begin{align}
	\mu^\circ_t = f(D - \Theta(\mu^\circ)) - f(-\Theta(\mu^\circ) - D). \label{eq:mueq} 
\end{align} 
We note that \eqref{eq:mueq} can be written as 
\begin{align}
\mu^\circ_t = 
	\begin{cases}
		f(D - \Theta(\mu^\circ)) &\mbox{ if } D >  \Theta(\mu^\circ), \\
		0 &\mbox{ if } |D| \leq \Theta(\mu^\circ), \\
		-f(-\Theta(\mu^\circ) - D) &\mbox{ if } D < -\Theta(\mu^\circ),
	\end{cases}
\end{align}
and thus, $D$ and $\mu^\circ_t$ always have the same sign. 
Moreover, $\mu^\circ$ is constant in time and the rod's response is purely viscoelastic (with natural curvature $\mu^\circ$) as long as $$|D| \leq \Theta(\mu^\circ).$$
In particular, the conditions imposed on the functions $f$ and $\Theta$ can be interpreted as the physical assumptions that if the rod is in its natural configuration, then there is a positive threshold for the magnitude of the driving force $|D|$ required to change its natural configuration, and this threshold grows as the magnitude of the natural curvature $|\mu^\circ|$ grows. In particular, if the driving force $D$ is positive (negative) and large enough in magnitude, then a positive (resp. negative) change in $\mu^\circ$ occurs. 

\section{Quasi-static Eulerian Strut} 

In this section we specialize our study of the general equations of motion to those modeling the quasistatic motion of an Eulerian strut. 

\subsection{Nondimensionalization}

Let $T$ and $F$ be the time and force scales used. We define the following dimensionless variables: 
\begin{gather}
	\bar s = \frac{1}{L} s, \quad \bar t = \frac{1}{T} t, \quad 
	\bar{\bs r}(\bar s, \bar t) = \frac{1}{L} \bs r(s,t), \quad \bar n(\bar s, \bar t) = \frac{1}{F} n(s,t), \\
	\bar \mu(\bar s, \bar t) = L \mu(s,t), \quad 
	\bar \mu^\circ(\bar t) = L \mu(t), \quad 
		\bar M(\bar s, \bar t) = \frac{L}{(EI)} M(s,t), \\
	\bar \kappa(\bar \mu^\circ) = \frac{L^2}{(EI)} \kappa(\mu^\circ), \quad 
	\bar D(\bar t) = \int_0^1 (\bar \mu^(\bar s, \bar t) - \bar \mu^\circ(\bar t)) d\bar s - \bar \kappa'(\bar \mu^\circ(t)), \\ 
	\bar \Theta(\bar \mu^\circ) = \frac{1}{(EI)} \Theta(\mu^\circ), \quad
	\bar f(x) = LT f((EI) \bar x). 
\end{gather}

Then the equations of motion are equivalent to: for all $(\bar s, \bar t) \in [0,1] \times [0,\infty)$, 
\begin{align}
\bar \eps_1 \bar{\bs r}_{\bar t \, \bar t} \cdot \bs a &= \bar N_{\bar s}, \\ 
\bar \eps_1 \bar{\bs r}_{\bar t \, \bar t} \cdot \bs b &= \bar H_{\bar s}, \\ 
\bar \eps_0 \theta_{\bar t \, \bar t} &= \bar \alpha \bar M_{\bar s} + \bar H, 
\end{align}
where (upon choosing appropriate units of time)
\begin{align}
\bar M = (\bar \mu - \bar \mu^\circ) + \bar{\mu}_{\bar t}, 
\end{align}
and $\bar \eps_0, \bar \eps_1, \bar \al$ 
are positive dimensionless constants. The evolution for the natural curvature is given by 
\begin{align}
\bar \mu_{\bar t}^{\circ} &= \bar f(\bar D - \bar \Theta(\bar \mu^\circ)) - 
\bar f(- \bar \Theta(\bar \mu^\circ) - \bar D). 
\end{align}
 Moreover, an assumption in rod theory is that a typical transversal length of the rod's cross section is much smaller than the length of the rod which implies that $\bar \eps_0 \ll \bar \eps_1$. In the remainder of this work, we will study the above system of equations in the quasi-static limit
\begin{align} 
	\bar \eps_0 = \bar \eps_1 = 0,
\end{align}
and we will drop the over bars on the remaining variables. 

\subsection{Eulerian strut} 

We consider the boundary conditions corresponding to an Eulerian strut: 
\begin{align}
	\bs r(0,t) = \bs o, \quad \theta(0,t) = 0, \quad \bs n(1,t) = -\beta \bs i, \quad M(1,t) = 0, \label{eq:boundary1}
\end{align}
where $\beta \geq 0$. 
Then $\bs n(s,t) = -\beta \bs i$ which implies that 
\begin{align}
	H(s,t) = \bs n(s,t) \cdot \bs b(s,t)
	= \beta \sin \theta(s,t) = \beta \sin \Bigl (
	\int_0^s \mu(\zeta,t) d\zeta
	\Bigr ). 
\end{align}
We have $\al M_s(s,t) = -H(s,t)$ which along with \eqref{eq:boundary1} and integration implies that $\al M(s,t) = \int_s^1 H(\sigma, t) d\sigma$. The equations of motion and boundary conditions reduce to the following coupled differential equations with initial conditions: for all $(s,t) \in [0,1] \times [0,T)$
\begin{align}
	&\mu_t(s,t) = - \mu(s,t) + \mu^\circ(t) + \gamma \int_s^1 \sin \Bigl (
	\int_0^\sigma \mu(\zeta,t) d\zeta
	\Bigr ) d\sigma, \label{eq:muequation} \\
	&\mu^\circ_t(t) = f\bigl (\hat D(\mu(\cdot,t), \mucir(t)) - \Theta(\mu^\circ(t))\bigr ) \\
	&\qquad \qquad - f\bigl (-\Theta(\mu^\circ(t)) - \hat D(\mu(\cdot,t) \bigr ), \label{eq:mucircequation} \\
	&\mu(s,0) = \mu_0(s), \quad \mu^\circ(0) = \mu^\circ_0, \label{eq:boundary}
\end{align}
where $\gamma = \beta/\al$ and $\hat D(\mu(\cdot,t), \mucir(t)) = \int_0^1 \mu(s,t) ds - \mu^\circ(t) - \kappa'(\mu^\circ(t))$. For $(\mu, \mucir) \in L^2(0,1) \times \bbR$, let  
\begin{align}
	 F(\mu, \mucir)(s) = \Bigl (&
	\mucir - \mu(s) + \gamma \int_s^1 \sin \Bigl ( \int_0^\sigma
	\mu(\zeta) d \zeta
	\Bigr )	d\sigma , \\
	&f\bigl (\hat D(\mu(\cdot,t), \mucir(t)) - \Theta(\mu^\circ(t))\bigr ) \label{eq:Fdef} \\
	&\qquad \qquad - f\bigl (-\Theta(\mu^\circ(t)) - \hat D(\mu(\cdot,t) \bigr )
	\Bigr ).
\end{align}
Then \eqref{eq:muequation}, \eqref{eq:mucircequation} and \eqref{eq:boundary} can be written as the ordinary differential equation in $L^2(0,1) \times \bbR$:   
\begin{align}
	&\p_t \bigl (
		\mu(\cdot,t),
		\mucir(t)
	\bigr )
= F(\mu(\cdot,t), \mucir(t)), \label{eq:evolution}\\
&(\mu(\cdot,0), \mucir(0)) = (\mu_0, \mu_0^\circ). 
\end{align}
It is straightforward to verify that $F : L^2(0,1) \times \bbR \rar L^2(0,1) \times \bbR$ is continuously Fr\'echet differentiable with Fr\'echet derivative $[DF(\mu, \mu^\circ)](\xi, \xi^\circ) =$ 
\begin{align}
 \Bigl (&
	\xi^\circ - \xi(s) + \gamma \int_s^1 \cos \Bigl ( \int_0^\sigma
\mu(\zeta) d \zeta
\Bigr )	\int_0^\sigma \xi(\zeta)d\zeta d\sigma , \\
&f\bigl (\hat D(\mu(\cdot,t), \mucir(t)) - \Theta(\mu^\circ(t))\bigr ) 
\Bigl [
\int_0^1 \xi(s) ds - \kappa''(\mu^\circ) \xi^\circ - \Theta'(\mu^\circ) \xi^\circ
\Bigr ] \label{eq:Frechet} \\
&+f\bigl (- \Theta(\mu^\circ(t)) - \hat D(\mu(\cdot,t), \mucir(t))\bigr ) 
\Bigl [
\int_0^1 \xi(s) ds - \kappa''(\mu^\circ) \xi^\circ + \Theta'(\mu^\circ) \xi^\circ
\Bigr ]
\Bigr ). 
\end{align}
Indeed, writing $\theta(\sigma) = \int_0^\sigma \mu(\zeta) d\zeta$ and $\phi(\sigma) = \int_0^\sigma \xi(\zeta) d\zeta$ we have that the first component of $F$ satisfies 
\begin{align}
F_1(\mu + \xi, \mucir + \xi^\circ) &= 
F_1(\mu, \mu^\circ) +
\xi^\circ - \xi(s) \\ &\quad + \gamma \int_s^1 [\sin (\theta(\sigma) + \phi(\sigma)) - \sin \theta(\sigma)] d\sigma.
\end{align}
By trigonometric identities and Cauchy-Schwarz we have 
\begin{align}
\Bigl | \int_s^1 & [\sin(\theta(\sigma) + \phi(\sigma)) - \sin \theta(\sigma)] d\sigma - \int_s^1 \cos \theta(\sigma) \phi(\sigma) d\sigma \Bigr | \\
&= \Bigl |\int_s^1 \sin \theta(\sigma) [\cos \phi(\sigma) - 1] d\sigma + 
\int_s^1 \cos \theta(\sigma) [\sin \phi(\sigma) - \phi(\sigma)] d\sigma \Bigr | \\
&\leq C ( \| \phi \|_{L^\infty(0,1)}^2 + \| \phi \|_{L^\infty(0,1)}^3 ) \\
&\leq C ( \| \xi \|_{L^2(0,1)}^2 + \| \xi \|_{L^2(0,1)}^3 )
\end{align}
where $C$ is an absolute constant. Thus, 
\begin{align}
	F_1(\mu + \xi, \mu^\circ + \xi^\circ) &- F_1(\mu, \mu^\circ) + 
	\Bigl (
	\xi^\circ - \xi(s) \gamma + \int_0^1 \phi(\sigma) \cos \theta(\sigma) d\sigma 
	\Bigr )
	\\&= o(\| \xi \|_{L^2(0,1)})
\end{align}
as $\| \xi \|_{L^2(0,1)} \rar 0$. Similarly, by using the smoothness assumptions for $\kappa$, $f$ and $\Theta$, we conclude that the second component of $F$ satisfies 
\begin{align}
	F_2&(\mu + \xi, \mu^\circ + \xi^\circ) = 
	F_2(\mu, \mu^\circ) \\
	&=f'\bigl (\hat D(\mu(\cdot,t), \mucir(t)) - \Theta(\mu^\circ(t))\bigr ) 
	\Bigl [
	\int_0^1 \xi(s) ds - \kappa''(\mu^\circ) \xi^\circ - \Theta'(\mu^\circ) \xi^\circ
	\Bigr ] \\
	&\quad+f'\bigl (- \Theta(\mu^\circ(t)) - \hat D(\mu(\cdot,t), \mucir(t)) \bigr ) 
	\Bigl [
	\int_0^1 \xi(s) ds - \kappa''(\mu^\circ) \xi^\circ + \Theta'(\mu^\circ) \xi^\circ
	\Bigr ] \\
	&\quad + o(\| \xi \|_{L^2(0,1)} + |\xi^\circ|)  
\end{align}
as $\| \xi \|_{L^2(0,1)} + |\xi^\circ| \rar 0$, proving \eqref{eq:Frechet}. 

\subsection{Global well-posedness for the quasistatic equations of motion} 

\begin{thm}\label{p:wellposedness}
Let $(\mu_0, \mucir_0) \in L^2(0,1) \times \bbR$.  Then there exists unique
$(\mu, \mu^\circ) \in C^1([0,\infty); L^2(0,1) \times \bbR)
$
	satisfying \eqref{eq:muequation}, \eqref{eq:mucircequation} and \eqref{eq:boundary}.
	
	 Moreover, there exists a constant $C(\mu_0, \mu_0^\circ, \gamma) > 0$ such that for all $t \in [0,\infty)$ 
	\begin{align}
	\| \mu(\cdot, t) \|_{L^2(0,1)} + |\mu^\circ(t)| \leq C. \label{eq:globalbound}
	\end{align}
\end{thm}

\begin{proof} 
Since \eqref{eq:muequation} and \eqref{eq:mucircequation} can be written as the differential equation \eqref{eq:evolution} with $F$ continuously differentiable on $L^2(0,1) \times \bbR$, 
the standard well-posedness theory for differential equations in Banach spaces implies that there exist $T_+ > 0$ and unique 
$(\mu, \mu^\circ) \in C^1([0,T_+); L^2(0,1) \times \bbR)$ solving \eqref{eq:muequation}, \eqref{eq:mucircequation} and \eqref{eq:boundary} on a maximal time of interval of existence $[0,T_+)$ (see e.g. Chapter 3 of \cite{Henry81Parabolic}). Moreover, we have the breakdown criterion  
\begin{align}
T_+ < \infty \implies
\limsup_{t \rar T_+} [ \| \mu(\cdot, t) \|_{L^2(0,1)} + |\mu^\circ(t)| ] = \infty, \label{eq:breakdown}
\end{align}
and we have continuous dependence on initial conditions: if 
\begin{align}
\lim_{n \rar \infty} \bigl \| (\mu_{0,n}, \mu^\circ_{0,n}) - (\mu_0, \mu^\circ_0) \bigr \|_{L^2(0,1) \times \bbR} = 0, 
\end{align}
then for all $n$ sufficiently large, the unique solution
$$(\mu_n, \mu_n^\circ) \in C^1([0,T_{+,n}); L^2(0,1) \times \bbR)
$$
to \eqref{eq:muequation} and \eqref{eq:mucircequation} with $(\mu_n(\cdot,0), \mu^\circ_n(0)) = (\mu_{0,n}, \mu^\circ_{0,n})$ exists on $[0,T_+)$, and for all $0 \leq T < T_+$, 
\begin{align}
\lim_{n \rar \infty} \sup_{t \in [0,T]} \bigl \|
(\mu_n(\cdot,t), \mu^\circ_n(t)) - (\mu(\cdot,t), \mu^\circ(t))
\bigr \|_{L^2(0,1) \times \bbR} = 0. \label{eq:dependence}
\end{align}

For $t \in [0,T_+)$ define 
\begin{align}
	V(\mu(\cdot,t), \mu^\circ(t)) &= \frac{1}{2} \int_0^1 (\mu(s,t) - \mu^\circ(t))^2 ds + \kappa(\mu^\circ(t)) \\
	&\quad + \gamma \int_0^1 \cos \theta(s,t) ds
\end{align}
where $\theta(s,t) = \int_0^s \mu(\zeta,t) d\zeta$.  We note that $V$ is the difference of the total energy stored by the rod and the total work done by the terminal thrust. Then \eqref{eq:muequation}, \eqref{eq:mucircequation}, the relation $\theta_{ts} = \theta_{st} = \mu_t$ and integration by parts imply that 
\begin{align}
	\frac{d}{dt} V(\mu(\cdot,t), \mu^\circ(t)) &= \int_0^1 (\mu(s,t) - \mu^\circ(t))\mu_t(s,t) ds \\ &\quad- \Bigl 
	[ \int_0^1 \mu(s,t) ds - \mu^\circ(t) - \kappa'(\mu^\circ(t)) \Bigr ]  \mu^\circ_t(t) \\
	&\quad - \gamma \int_0^1 \sin \theta(s,t) \theta_t(s,t) ds \\
	&= - \int_0^1 \mu_t^2(s,t) ds + \gamma \int_0^1 \Bigl [
	\int_s^1 \sin \theta(\sigma,t) d\sigma \Bigr 
	] \theta_{st}(s,t) ds \\
	&\quad - \hat D(\mu(\cdot,t), \mu^\circ(t)) \mu_t^\circ(t) 
	- \gamma \int_0^1 \sin \theta(s,t) \theta_t(s,t) ds \\
	&= -\int_0^1 \mu_t^2(s,t) ds - |\hat D(\mu(\cdot,t), \mu^\circ(t)) \mu_t^\circ(t)|  \label{eq:liapunov} 
\end{align}
where the last equality follows from the fact that $\hat D(\mu(\cdot,t), \mu^\circ(t))$ and $\mu_t^\circ(t)$ have the same sign via \eqref{eq:mucircequation}. 
Thus, for all $t \in [0,T_+)$ 
\begin{align}
\frac{1}{2} \int_0^1 (\mu(s,t) - \mu^\circ(t))^2 ds &+ \kappa(\mu^\circ(t)) \\
&\leq V(\mu(\cdot,t), \mu^\circ(t)) - \gamma \int_0^1 \cos \theta(s,t) ds \\
&\leq V(\mu_0, \mu^\circ_0) + \gamma. 
\end{align}
By our assumptions on $\kappa$ we conclude that $\mu^\circ : [0,T_+) \rar \bbR$ is bounded whence $\mu(\cdot,t) : [0,T_+) \rar L^2(0,1)$ is bounded as well. By \eqref{eq:breakdown} we conclude that $T_+ = \infty$ as well as \eqref{eq:globalbound}. 
\end{proof}

\subsection{Equilibrium points}

The equilibrium points $(\nu, \nu^\circ) \in L^2(0,1) \times \bbR$ of the evolution equation \eqref{eq:evolution} satisfy $F(\nu, \nu^\circ) = (0,0)$ i.e. 
\begin{gather}
\nu^\circ - \nu(s) + \gamma \int_s^1 \sin \Bigl ( \int_0^\sigma 
\nu(\zeta) d\zeta
\Bigr ) d\sigma = 0, \quad s \in [0,1], \label{eq:equ1}\\
|\hat D(\nu, \nu^\circ)| \leq \Theta(\nu^\circ). \label{eq:equ2} 
\end{gather}
We observe that \eqref{eq:equ1} implies that $\nu \in C^1([0,1])$ and $\nu(1) = \nu^\circ$. Defining $\phi(s) = \int_0^s \nu(\zeta) d\zeta \in C^2([0,1])$, we see that 
\begin{align}
(\nu, \nu^\circ) = (\phi', \phi'(1)),
\end{align}
and 
\eqref{eq:equ1}, \eqref{eq:equ2} are equivalent to 
\begin{align}
	&\phi''(s) + \gamma \sin \phi(s) = 0, \quad s \in [0,1], \label{eq:phieq}
\end{align}
$\phi(0) = 0, \phi'(1) = \nu^\circ,$ and 
\begin{align}
	\Bigl | \phi(1) - \phi'(1) - \kappa'(\phi'(1)) \Bigr |\leq \Theta(\phi'(1)). \label{eq:nucirceq}
\end{align}

The trivial equilibrium point 
\begin{align}
	\nu = 0, \quad \nucir = 0 
\end{align}
gives the configuration of a straight rod. 
 We note that since $\hat D(0,0) = 0$ and $\Theta$ is bounded from below there always exist nontrivial equilibrium points. Indeed, by continuous dependence on initial conditions for \eqref{eq:phieq}, there exists $\delta > 0$ such that for all $\alpha \in (-\delta, \delta)$ the unique solution $\phi_\al \in C^2([0,1])$ to \eqref{eq:phieq} satisfying the initial conditions $\phi_\al(0) = 0, \phi_\al'(0) = \al$ also satisfies 
\begin{align}
	|\hat D(\phi'_\al, \phi_\al'(1))| < \min_{\nu^\circ \in \bbR} \Theta(\nu^\circ) \leq \Theta(\phi_\al'(1)).
\end{align}
Then $(\phi_\al', \phi'_\al(1))$ is a nontrivial equilibrium point of \eqref{eq:evolution} for $\al \in (-\delta, \delta) \backslash \{0\}$. 

We now turn to the stability properties of equilibrium points, and in what follows, $(\nu, \nu^\circ) \in C^1([0,1]) \times \bbR$ is a fixed equilibrium point and $\phi(s) = \int_0^s \nu(\zeta) d\zeta$. The linearization of \eqref{eq:evolution} about $(\nu, \nu^\circ)$ is given by 
\begin{align}
\p_t (
		\mu(s, t),
		\mu^\circ(t)
	)
&= [DF(\nu, \nu^\circ)](
	\mu(s, t),
	\mu^\circ(t)) \\
&= 
\Bigl (
\mu^\circ(t) - \mu(s,t) + \int_s^1 \int_0^\sigma \mu(\zeta,t) \cos \phi(\sigma) d\zeta d\sigma,
0 
\Bigr ) \\
&= 
\Bigl (
	\mu^\circ(t) - \mu(s,t) + \int_0^1 K(s,\zeta) \mu(\zeta,t) d\zeta,
	0 \Bigr )
\end{align}
where $K(s,\zeta) = \int^1_{\max(s,\zeta)} \cos \phi(\sigma) d\sigma$. We denote $\cl L = DF(\nu, \nu^\circ) : L^2(0,1) \times \bbR \rar L^2(0,1) \times \bbR$. Since $\cl L$ is bounded, the solution to the linearized equations about $(\nu, \nu^\circ)$ can be written succinctly as  
\begin{align}
(
		\mu(\cdot, t),
		\mu^\circ(t)
	)
= e^{t \cl L} (
	\mu(\cdot,0),
	\mucir(0)). \label{eq:linearizedaboutequ}
\end{align} 
The next proposition shows that there are only finitely many unstable directions when perturbing $(\nu, \nu^\circ)$ (at the linearized level) and 0 is \emph{always} a simple eigenvalue for $\cl L$. 

\begin{prop}\label{p:spectrum}
The spectrum 
\begin{align}
\sigma(\cl L) = \{ \lambda \in \bbC \mid \cl L - \lambda I \mbox{ does not have a bounded inverse }\}
\end{align}
satisfies 
\begin{enumerate}
	\item $\sigma(\cl L) \subset \bbR$ is countable, \\
	\item $-1 \in \sigma(\cl L)$ and $-1$ is the only possible limit point of $\sigma(\cl L)$, \\
	\item every $\lambda \in \sigma(\cl L) \backslash \{-1,0\}$ is an eigenvalue with associated one dimensional eigenspace, \\
	\item $0 \in \sigma(\cl L)$ and
	\begin{align}
		\ker \cl L = \mbox{span} (\theta_0', \theta_0'(1))
	\end{align}  
where $\theta_0 \in C^2([0,1])$ is the unique solution to 
\begin{align}
	&\theta''(s) + \gamma [\cos \phi(s)] \theta(s) = 0, \\
	&\theta(0) = 0, \quad \theta'(0) = 1. \label{eq:thetazero}
\end{align}
\end{enumerate}
\end{prop}

\begin{proof}

We first consider the invertibility properties of $\cl L - \lambda I : L^2(0,1) \times \bbR \rar L^2(0,1) \times \bbR$ for $\lambda \in \bbC \backslash \{0\}$. For $(\xi, \xi^\circ) \in L^2(0,1) \times \bbR$, we have $(\cl L - \lambda I)(\mu, \mu^\circ) = (\xi, \xi^\circ)$ if and only if $\mu^\circ = -\frac{1}{\lambda} \xi^\circ$ and 
\begin{align}
	-(1 + \lambda) \mu(s) + \int_0^1 K(s,\zeta) \mu(\zeta) d\zeta = 
	\xi(s) + \frac{1}{\lambda} \xi^\circ. \label{eq:IKeq}
\end{align} 
Denote the integral operator $\cl K \mu(s) = \int_0^1 K(s,\zeta) \mu(s) ds$. Then \eqref{eq:IKeq} is solvable for arbitrary $(\xi, \xi^\circ) \in L^2(0,1) \times \bbR$ if and only if $-(1+\lambda)I + \cl K : L^2(0,1) \rar L^2(0,1)$ has a bounded inverse. Thus, $\lambda \in \sigma(\cl L) \backslash 0$ if and only if $1+\lambda \in \sigma(\cl K) \backslash \{1\}$. Since $K(s,\zeta) = K(\zeta,s) = \overline{K(\zeta,s)}$ and $K \in C([0,1] \times [0,1])$, $\cl K$ is a self-adjoint compact operator. By the spectral theorem for compact self-adjoint operators, we can immediately conclude $(1)$, $(2)$ and $\lambda \in \sigma(\cl L) \backslash\{-1,0\}$ if and only if there exists $\mu \in L^2(0,1) \backslash \{0\}$ satisfying
\begin{align}
	-(1 + \lambda) \mu(s) + \int_0^1 K(s,\zeta) \mu(\zeta) = 0. \label{eq:Kprop}
\end{align} 
But for $\lambda \in \bbC \backslash \{0\}$, $(\cl L - \lambda I)(\mu, \mu^\circ) = (0,0)$ if and only if $\mu^\circ = 0$ and $\mu$ satisfies \eqref{eq:Kprop}. Thus, every $\lambda \in \sigma(\cl L) \backslash \{-1,0\}$ is an eigenvalue. 

Let $\lambda \in \sigma(\cl L) \backslash \{-1,0\}$. We claim that $\dim \ker (\cl L - \lambda I) = 1$. Let $\theta_\la, \psi_\la \in C^2([0,1])$ be the fundamental system for the linear differential equation
\begin{align}
&(1+\lambda) \theta''(s) + \gamma [\cos \phi(s)] \theta(s) = 0, \label{eq:fundsoln}
\end{align} 
satisfying 
\begin{align}
\theta_\la(0) = 0, \quad \theta_\la'(0) = 1, \\
\psi_\la(0) = 1, \quad \psi'_\la(0) = 0. 
\end{align}
By linearity, every solution to \eqref{eq:fundsoln} is a linear combination of $\theta_\la$ and $\psi_\la$.  
If $(\mu, \mu^\circ) \in L^2(0,1) \times \bbR$ satisfies $(\cl L - \lambda I)(\mu, \mu^\circ) = (0,0)$, then $\mu^\circ = 0$ and $\mu$ satisfies \eqref{eq:Kprop}. Thus, $\mu \in C^1([0,1])$ and $\theta(s) = \int_0^s \mu(\sigma) d\sigma$ verifies \eqref{eq:fundsoln} (by differentiating \eqref{eq:Kprop}) and $\theta(0) = 0$. Thus, $\theta = \theta'(0)\theta_\lambda$ so $\mu = \theta'(0) \theta'_\lambda$ and $\ker (\cl L - \lambda I) \subset \mbox{span}\, (\theta_\lambda',\theta_\lambda'(1))$. This proves the claim (and thus, $(3)$). 

What remains is to prove $(4)$. Let $\theta_0 \in C^2([0,1])$ be the solution to \eqref{eq:thetazero}. Then $(\mu_0, \mu_0^\circ) = (\theta_0', \theta_0'(1))$ verifies 
\begin{align}
	-\mu_0(s) + \mu_0^\circ + \int_0^1 K(s,\zeta) \mu_0(\zeta) d\zeta = 0,
\end{align}
which proves $\ker \cl L \not = \{0\}$. The proof that $\ker \cl L = \mbox{span}(\theta_0', \theta_0'(1))$ is nearly identical to that showing $\ker(\cl L - \lambda I) = \mbox{span}(\theta_\lambda', \theta_\lambda'(1))$ for $\lambda \notin \sigma(\cl L) \backslash \{-1,0\}$ and is omitted.  
\end{proof}

The following proposition elucidates the underlying source of the kernel of $\cl L$ and provides a simple orbital asymptotic stability statement. 

\begin{prop}\label{p:curve}	Assume that 
	$|\hat D(\nu, \nu^\circ)| < \Theta(\nu^\circ)$, and let $\theta_0$ be as in Proposition \ref{p:spectrum}.
	\begin{enumerate}
\item There exist $\eps > 0$ and a smooth embedded curve 
		\begin{align}
			(-\eps, \eps) \ni \alpha \mapsto (\nu(\alpha), \nu^\circ(\alpha)) \in L^2(0,1) \times \bbR
		\end{align}
	 such that $(\nu(0),\nu^\circ(0)) = (\nu, \nu^\circ)$, for all $\al \in (-\eps, \eps)$, $(\nu(\al),\nucir(\al))$ is an equilibrium point of \eqref{eq:evolution}, and $\frac{d}{d\al}(\nu(\al), \nucir(\al)) |_{\al = 0} = (\theta_0', \theta'_0(1))$.
	 
	 \item If $\sigma(\cl L) \cap \{ \lambda \mid \Re \lambda > 0 \} = \varnothing$, then there exist $\delta > 0$, $C >  0$ and $\beta > 0$ such that if 
	 \begin{align}
	 	 \| (\mu_0, \mu_0^\circ) - (\nu(0), \nu^\circ(0)) \|_{L^2(0,1) \times \bbR} < \delta, 
	 \end{align}
 then there exists $\al_\infty \in (-\eps,\eps)$ such that the solution $(\mu, \mu^\circ) \in$ \newline $C^1([0,\infty); L^2(0,1) \times \bbR)$ to \eqref{eq:evolution} with initial data $(\mu_0, \mu_0^\circ)$ satisfies 
 \begin{align}
 	\| (\mu(\cdot,t),\mu^\circ(t)) - (\nu(\al_\infty), \nu^\circ(\al_\infty)) \|_{L^2(0,1) \times \bbR} \leq C \delta e^{-\beta t}.
 \end{align}
\end{enumerate}
\end{prop}

\begin{proof} (1)
	For $\alpha \in \bbR$, let $\phi_\alpha \in C^2([0,1])$ be the unique solution to 
	\begin{align}
		&\phi_\alpha''(s) + \gamma \sin \phi_\alpha(s) = 0, \\ 
		&\phi_\alpha(0) = 0, \quad \phi_\alpha'(0) = \phi'(0) + \alpha. \label{eq:phieps}
	\end{align}
We note that $\phi_0(s) = \phi(s) = \int_0^s \nu(\sigma) d\sigma$. Since $|\hat D(\nu, \nu^\circ)| < \Theta(\nu^\circ)$, continuous dependence on initial conditions implies that there exists $\eps > 0$ such that for all $\alpha \in (-\eps, \eps)$, 
\begin{align}
	|\hat D(\phi_\alpha', \phi_\alpha'(1))| < \Theta(\phi'_\alpha(1)).  
\end{align}
Thus, $(\nu(\al),\nu^\circ(\al)) = (\phi'_\al, \phi'_\alpha(1)) : (-\eps, \eps) \rar L^2(0,1) \times \bbR$ is a smooth embedded curve consisting entirely of equilibrium points of \eqref{eq:evolution}. Finally, by \eqref{eq:phieps} we see that $\dot \phi = \frac{\p \phi}{\p \alpha} |_{\alpha = 0}$ verifies 
\begin{align}
		&\dot \phi''(s) + \gamma [\cos \phi(s)] \dot \phi(s) = 0, \\ 
		&\dot \phi(0) = 0, \quad \dot \phi'(0) = 1,
\end{align}
and thus, $\dot \phi = \theta_0$ so $\frac{d}{d\alpha}(\nu(\alpha), \nu(\alpha)^\circ) |_{\al = 0} = (\theta'_0, \theta'_0(1)).$

(2) The statement follows from (1) and a general asymptotic stability result discussed on p. 108-109 of \cite{Henry81Parabolic}. For completeness, we will briefly sketch the details. By redefining $F$ we may assume without loss of generality that $(\nu, \nu^\circ) = (0,0)$, and rewrite \eqref{eq:evolution} as 
\begin{align}
	\p_t(\mu(\cdot,t), \mu^\circ(t)) = \cl L(\mu(\cdot,t), \mu^\circ) + G(\mu(\cdot, t), \mu^\circ(t)), \label{eq:evol2}
\end{align} 
where $\cl L = DF(0,0)$ and $G(\mu, \mu^\circ) = F(\mu, \mu^\circ) - \cl L(\mu, \mu^\circ)$ is continuously Fr\'echet differentiable and satisfies $G(0,0) = (0,0)$ and $DG(0,0) = 0$ (as operators). We decompose 
$$L^2(0,1) \times \bbR = \spn(\theta_0',\theta_0'(1)) \oplus \cl H$$ where $\cl H$ is the image of the spectral projection $\Pi : L^2(0,1) \times \bbR \rar L^2(0,1) \times \bbR$ associated to $\sigma(\cl L) \backslash \{0\}$. We then write $$(\mu(\cdot,t), \mu^\circ(t)) = (\nu(\al(t)), \nu^\circ(\al(t))) +(\xi(\cdot,t), \xi^\circ(t)),$$ where $(\xi,\xi^\circ) \in \cl H$, and insert this ansatz into \eqref{eq:evol2}, obtaining evolution equations for $\al$ and $(\xi,\xi^\circ)$.  Since $G(0,0) = (0,0)$, $DG(0,0) = 0$ (as operators) and $\cl L |_{\cl H}$ has spectrum contained in a compact subset of $(-\infty,0)$, Duhamel's principle applied to the equation satisfied by $(\xi,\xi^\circ)$) implies that if $\sup_{t \in [0,T]} |\al(t)| < 2\delta$ with $\delta$ small, then for all $t \in [0,T]$,
\begin{align}
	\| (\xi(\cdot,t), \xi^\circ(t)) \|_{L^2(0,1) \times \bbR} \leq C_1 e^{-\beta t} \| (\xi(\cdot,0), \xi^\circ(0)) \|_{L^2(0,1) \times \bbR}, 
\end{align}
where $C_1$ is an absolute constant. This then implies via the equation satisfied by $\al$ that $|\p_t \al(t)| \leq C_2 e^{-\beta t} \| (\xi(\cdot,0), \xi^\circ(0)) \|_{L^2(0,1) \times \bbR}$ where $C_2$ is an absolute constant. A bootstrap argument then concludes the proof. 
\end{proof}

We remark that Proposition \ref{p:curve} (1) generalizes our observation at the start of this subsection that there always exist nontrivial equilibrium points, and Proposition \ref{p:curve} (2) applies to $(\nu,\nu^\circ) = (0,0)$ as long as $\gamma \leq \frac{\pi^2}{4}$, the critical value for buckling of the straight rod. An interesting question is the existence and local structure (in $L^2(0,1) \times \bbR$) of the set of limiting equilibrium points satisfying
\begin{align}
	\Bigl |\int_0^1 \nu(s) ds - \nu^\circ - \kappa'(\nu^\circ) \Bigr | = \Theta(\nu^\circ). 
\end{align}  
The answer should be highly dependent on the specific choices of $\kappa$ and $\Theta$ and will not be addressed here.   

\section{Quasi-static Eulerian Strut: Asymptotics} 

In this final section, we prove that every solution $(\mu(\cdot,t), \mu^\circ(t))$ to \eqref{eq:evolution} converges to an equilibrium point of \eqref{eq:evolution} in $L^2(0,1) \times \bbR$ as $t \rar \infty$. 

\subsection{Liapunov function}

In the proof of Proposition \ref{p:wellposedness} we used that 
\begin{align}
	V(\mu(\cdot,t), \mu^\circ(t)) &= \frac{1}{2} \int_0^1 (\mu(s,t) - \mu^\circ(t))^2 ds + \kappa(\mu^\circ(t)) \\
	&\quad + \gamma \int_0^1 \cos \theta(s,t) ds,
\end{align}
where $\theta(s,t) = \int_0^s \mu(\sigma,t) d\sigma$, is a Liapunov function: 
\begin{align}
	\frac{d}{dt} V(\mu(\cdot,t), \mu^\circ(t)) &= 
-\int_0^1 \mu_t^2(s,t) ds - |\hat D(\mu(\cdot,t), \mu^\circ(t))|  |\mu_t^\circ(t)| \label{eq:liap2} \\
&\leq 0. 
\end{align}

\begin{lem}\label{l:liapunov}
	Let $(\nu, \nucir) \in C^1([0,\infty); L^2(0,1) \times \bbR)$ satisfy \eqref{eq:evolution}. If there exists $T > 0$ such that for all $t \in [0,T]$
	\begin{align}
		\int_0^1 \nu_t^2(s,t) ds + |\hat D(\nu(\cdot,t), \nu^\circ(t))||\nucir_t(t)| = 0, \label{eq:liapass}
	\end{align}
	then $(\nu(\cdot,t), \nu^\circ(t))$ is constant in time, and $(\nu(\cdot,0), \nu^\circ(0))$ is an equilibrium point of \eqref{eq:evolution}.  
\end{lem}

\begin{proof}
It is clear from \eqref{eq:liapass} that our assumptions immediately imply that $\nu_t(\cdot,t) = 0$ for all $t \in [0,T]$. Suppose that there exists a time $t_0 \in [0,T]$ such that  
\begin{align}
	|\hat D(\nu(\cdot,t_0), \nu^\circ(t_0))||\nucir_t(t_0)| = 0
\end{align}
but $\nucir_t(t_0) \neq 0$. Then $|\hat D(\nu(\cdot,t_0), \nu^\circ(t_0))| = 0$ which by \eqref{eq:mucircequation} implies that 
$\nucir_t(t_0) = 0$, a contradiction. Thus, $\nu_t(t) = 0$ for all $t \in [0,T]$. By \eqref{eq:evolution} we conclude that $(\nu(\cdot,0), \nucir(0))$ is an equilibrium point of \eqref{eq:evolution}, so by uniqueness of solutions to \eqref{eq:evolution}, $(\nu(\cdot,t), \nucir(t)) = (\nu(\cdot,0), \nucir(0))$ for all $t$. 
\end{proof}

\subsection{Convergence to an equilibrium point} 

Our first step towards proving the convergence of a solution to \eqref{eq:evolution} to an equilibrium point of \eqref{eq:evolution} is proving the precompactness of the trajectory in $L^2(0,1) \times \bbR$.  

\begin{lem}\label{l:compact}
	Let $(\mu, \mucir) \in C^1([0,\infty); L^2(0,1) \times \bbR)$ solve \eqref{eq:evolution}. Then
	\begin{align}
		K = \{ (\mu(\cdot,t), \mu^\circ(t)) \mid t \in [0,\infty) \} 
	\end{align}
is precompact in $L^2(0,1) \times \bbR$. 
\end{lem}

\begin{proof}
By \eqref{eq:globalbound} it follows that $\{ \mu^\circ(t) \mid t \in [0,\infty)\}$ is precompact in $\bbR$. Let 
\begin{align}
	\xi(s,t) = \int_0^t e^{-(t-\tau)} \mu^\circ(\tau) d \tau +
	\gamma \int_0^t \int_s^1 e^{-(t-\tau)} \sin \theta(\sigma, \tau) d \sigma d\tau
\end{align}
where, as before, $\theta(s,t) = \int_0^s \mu(\zeta,t) d\zeta$. Then \eqref{eq:muequation} implies that for all $(s,t)$
\begin{align}
	\mu(s,t) = e^{-t} \mu_0(s) + \xi(s,t),
\end{align}
so it suffices to show that $\{ \xi(\cdot,t) \mid t \in [0,\infty)\}$ is precompact in $L^2(0,1)$. 

By \eqref{eq:globalbound} and repeated use of the triangle inequality we deduce that for all $(s,t)$ 
\begin{align}
|\xi(s,t)| \leq \int_0^t e^{-(t-\tau)} C + \gamma \int_0^t \int_s^1 e^{-(t-\tau)} d\sigma d\tau \leq C + \gamma. 
\end{align}
For all $0 \leq s_1 \leq s_2 \leq 1$ and $t$ we also obtain 
\begin{align}
|\xi(s_2,t) - \xi(s_1,t)| &= \Bigl |
\gamma \int_0^t \int_{s_1}^{s_2} e^{-(t-\tau)} \sin \theta(\sigma, \tau) d \sigma d\tau \Bigr |\\
&\leq \gamma \int_0^t \int_{s_1}^{s_2} e^{-(t-\tau)} d\sigma d\tau 
= \gamma|s_2 - s_1|. 
\end{align}
By the Arzela-Ascoli theorem we conclude that $\{ \xi(\cdot,t) \mid t \in [0,\infty) \}$ is precompact in $C([0,1]) \subset L^2(0,1)$ as desired. 
\end{proof}

Let $(\mu_0, \mucir_0) \in L^2(0,1) \times \bbR$, and let $(\mu, \mucir) \in C^1([0,\infty); L^2(0,1) \times \bbR)$ be the unique solution to \eqref{eq:muequation} and \eqref{eq:mucircequation} with $(\mu(\cdot,0), \mucir(0)) = (\mu_0, \mucir_0)$. We denote the evolution operator by 
$S(t)(\mu_0, \mucir_0)$ so that 
\begin{align}
	S(t)(\mu_0, \mucir_0) = (\mu(\cdot,t), \mu^\circ(t)). 
\end{align}
We recall the notion of the $\omega$-limit set relative to the semi-flow $S(t)$: $(\nu, \nucir) \in \omega(\mu_0, \mucir_0)$ if and only if there exists a sequence $t_n \geq 0$ such that $t_n \rar \infty$ and $S(t_n)(\mu_0, \mucir_0) \rar (\nu, \nucir)$ in $L^2(0,1) \times \bbR$. We now show that the $\omega$-limit set consists of equilibrium points.  

\begin{lem}\label{l:om}
	For all $(\mu_0, \mucir_0) \in L^2(0,1) \times \bbR$, $\omega(\mu_0, \mucir_0)$ is a nonempty subset of equilibrium points of \eqref{eq:evolution}. 
\end{lem}

\begin{proof}
By Lemma \ref{l:compact} it follows that $\omega(\mu_0, \mucir_0) \not =  \varnothing$. 
Let $(\nu_0, \nucir_0) \in \om(\mu_0, \mucir_0)$. Then there exists $t_n \rar \infty$ such that 
\begin{align}
	(\mu(\cdot,t_n), \mucir(t_n)) = S(t_n)(\mu_0, \mucir_0) \rar (\nu_0, \nucir_0). 
\end{align}
By continuous dependence on initial conditions \eqref{eq:dependence}, for all $T \in [0,\infty)$, 
\begin{align}
\lim_{n \rar \infty}	\sup_{t \in [0,T]} \left \| (\mu(\cdot, t_n + t), \mu^\circ(t_n+t)) - S(t)(\nu_0, \nu^\circ_0) \right \|_{L^2(0,1) \times \bbR} = 0. \label{eq:conv} 
\end{align}
Let $(\nu(\cdot,t), \nu^\circ(t)) = S(t)(\nu_0, \nu_0^\circ)$. 
We now show that $(\nu_0, \nucir_0)$ is an equilibrium point of \eqref{eq:evolution} i.e. $(\nu(\cdot,t), \nu^\circ(t)) = (\nu_0, \nucir_0)$ for all $t$.  By Lemma \ref{l:liapunov} $V(\mu(\cdot,t), \mucir(t))$ is non increasing and bounded below. Thus, $$\ell = \lim_{t \rar \infty} V(\mu(\cdot,t), \mucir(t))$$ exists. Let $t \in [0,\infty)$. By \eqref{eq:conv}, for all $s \in [0,1]$
\begin{align}
\lim_{n \rar\infty} \int_0^s \mu(\sigma, t_n + t) d\sigma =
\int_0^s \nu(\sigma,t) d\sigma. 
\end{align}
By the dominated convergence theorem we conclude that 
\begin{align}
	\lim_{n \rar \infty} \int_0^1 \cos \theta(s,t_n + t) ds &= 
	\lim_{n \rar \infty} \int_0^1 \cos \Bigl (
	\int_0^s \mu(\sigma, t_n + t) d\sigma 
	\Bigr ) ds \\&= \int_0^1 \cos \Bigl (
	\int_0^s \nu(\sigma,t) d\sigma 
	\Bigr ). 
\end{align}
By \eqref{eq:conv} we also have 
\begin{align}
	\lim_{n \rar \infty} \frac{1}{2} \int_0^1 (\mu(s,t_n + t&) - \mucir(t_n + t))^2 ds + \kappa(\mucir(t_n + t))\\ &= 
	\frac{1}{2} \int_0^1 (\nu(s, t) - \nucir(t))^2 ds + \kappa(\nucir(t)). 
\end{align}
Thus, 
\begin{align}
	\ell = \lim_{n \rar \infty}  V(\mu(\cdot, t_n + t), \mucir(t_n + t)) = V(\nu(\cdot,t), \nucir(t)), 
\end{align}
for all $t \in [0,\infty)$. By \eqref{eq:liap2} and Lemma \ref{l:liapunov} we conclude that $\nu_t(\cdot,t) = 0$ and $\nucir_t(t) = 0$ for all $t \in [0,\infty)$ as desired. 
\end{proof}

We will use compactness of the trajectory, the spectral information from Section 3 (Proposition \ref{p:spectrum}) and the following convergence theorem for dynamical systems to finish the proof of our convergence result for solutions to \eqref{eq:evolution}. The general set-up is as follows.
 
Let $X$ be a Banach space and $F \in C(X;X)$. Let $x \in X$, and let $\omega_F(x)$ denote the $\omega$-limit set of $x$ relative to $F$: $z \in \om_F(x)$ if and only if there exists $n_k \rar \infty$ such that $F^{n_k}(x) \rar z$. The set of fixed points of $F$ is denoted $\mbox{Fix}(F) = \{ y \in X \mid F(y) = y \}$. We assume the following hypotheses: 
\begin{itemize}
	\item $\omega_F(x) \subseteq \mbox{Fix}(F)$,
	\item $y \in \omega_F(x)$ and there exists a neighborhood $U$ of $y$ such that $F|_{U} : U \rar X$ is continuously Fr\'echet differentiable, 
	\item $\sigma (DF(y)) = \sigma^u \cup \sigma^c \cup \sigma^s$ where, $\sigma^u$, $\sigma^c$, and $\sigma^s$ are closed subsets of $\{ \lambda \in \bbC \mid |\lambda| > 1 \}$, $\{ \lambda \in \bbC \mid |\lambda| = 1 \}$, and $\{\lambda \in \bbC \mid |\lambda| < 1 \}$ respectively.
\end{itemize}
Let $X^i$ be the image of the spectral projection of $DF(y)$ associated with the spectral set $\sigma^i$, $i = u,c,s$. Brunovsk\'y and Pol\'acik proved the following in \cite{BPolacik97}.

\begin{thm}[Theorem B \cite{BPolacik97}]\label{t:bp97}
Assume that either $X^u$ is finite-dimensional or the trajectory of $x$, $\{ F^n(x) \}_{n = 0}^\infty$, is precompact in $X$ and one of the following properties hold: 
\begin{enumerate}[(a)]
\item $\dim X^c = 1$ and the trajectory is precompact, 
\item $\dim X^c = m < \infty$ and there is a submanifold $M \subseteq X$ with $\dim M = m$ such that $y \in M \subseteq \mbox{Fix}(F)$
\end{enumerate} 
Then $\omega_F(x) = \{ y\}$. 
\end{thm}

	We remark that the proof of Theorem \ref{t:bp97} essentially reduces to the case of assuming $(b)$. Assuming $(b)$ and $\omega_F(x) \neq \{y\}$, one can prove, using invariant manifold theory, that the local center manifold $W^c_{loc} = M$, and $\omega_F(x)$ contains a point distinct from $y$ in the local (strong) unstable manifold $W^u_{loc}$ (see Theorem A of \cite{BPolacik97}). This contradicts the fact that $\omega_F(x) \subseteq \mbox{Fix}(F)$, and thus, $\omega_F(x) = \{y\}$.

\begin{thm}
	Let $(\mu_0, \mu_0^\circ) \in L^2(0,1) \times \bbR$. Then there exists an equilibrium point $(\nu, \nu^\circ) \in L^2(0,1) \times \bbR$ of \eqref{eq:evolution} such that $\omega(\mu_0, \mu_0^\circ) = \{ (\nu, \nu_0)\}$, i.e. 
	\begin{align}
		\lim_{t \rar \infty} S(t)(\mu_0, \mu_0^\circ) = (\nu, \nu^\circ). 
	\end{align}
\end{thm}

\begin{proof}
We will apply Theorem \ref{t:bp97} in conjunction with Proposition \ref{p:spectrum} and Lemma \ref{l:compact}. Consider the time-1 map: 
\begin{align}
	\Phi(\mu_0, \mu_0^\circ) = S(1)(\mu_0, \mu_0^\circ), \quad
	(\mu_0, \mu_0^\circ) \in L^2(0,1) \times \bbR. \label{eq:phiS}
\end{align}
Let $\omega_{\Phi}(\mu_0, \mu_0^\circ)$ denote the $\omega$-limit set of the trajectory $\{ \Phi^n(\mu_0, \mu_0^\circ) \}_{n = 0}^\infty$. By \eqref{eq:phiS} and the semigroup property of $S(t)$, we clearly have $\om_{\Phi}(\mu_0, \mu_0^\circ) \subseteq \omega(\mu_0, \mu_0^\circ)$. We claim that $\om_{\Phi}(\mu_0, \mu_0^\circ) \supseteq \omega(\mu_0, \mu_0^\circ)$ and thus, 
\begin{align}
	\om_{\Phi}(\mu_0, \mu_0^\circ) = \omega(\mu_0, \mu_0^\circ) \label{eq:omequal}. 
\end{align}
Suppose that there exist $n_k \in \bbN$ and $t_{k} \in [n_k-1, n_{k}]$ such that $t_k \rar \infty$ and 
\begin{align}
	S(t_{k})(\mu_0, \mu_0^\circ) \rar (\nu, \nu^\circ). 
\end{align}
By Lemma \ref{l:om}, $(\nu, \nu^\circ)$ is an equilibrium point of \eqref{eq:evolution}. By continuous dependence on initial conditions \eqref{eq:dependence}, we have
\begin{align}
	0 &= \lim_{k \rar \infty} \sup_{t \in [0,1]} \bigl \|
	S(t)S(t_{k})(\mu_0, \mu_0^\circ) - S(t)(\nu, \nu^\circ) 
	\bigr \|_{L^2(0,1) \times \bbR} \\
	&= \lim_{k \rar \infty} \sup_{t \in [0,1]} \bigl \|
	S(t_{k}+t)(\mu_0, \mu_0^\circ) - (\nu, \nu^\circ) 
	\bigr \|_{L^2(0,1) \times \bbR}. 
\end{align}
In particular, we conclude that 
\begin{align}
0 &= \lim_{k \rar \infty} \bigl \|
S(n_k)(\mu_0, \mu_0^\circ) - (\nu, \nu^\circ) 
\bigr \|_{L^2(0,1) \times \bbR} \\
&= \lim_{k \rar \infty} \bigl \| 
\Phi^{n_k}(\mu_0, \mu_0^\circ) - (\nu, \nu^\circ) 
\bigr \|_{L^2(0,1) \times \bbR},
\end{align}
and thus, $(\nu, \nu^\circ) \in \om_{\Phi}(\mu_0, \mu^\circ_0)$, proving \eqref{eq:omequal}. 

By Theorem 3.4.4 of \cite{Henry81Parabolic}) we have that $\Phi : L^2(0,1) \times \bbR \rar L^2(0,1) \times \bbR$ is continuously Fr\'echet differentiable with 
\begin{align}
	[D\Phi(\mu_0, \mu_0^\circ)](\xi_0, \xi_0^\circ) = (\xi_L(\cdot,1), \xi_L^\circ(1)), 
\end{align}
where $(\xi_L, \xi_L^\circ) \in C^1([0,\infty); L^2(0,1) \times \bbR)$ solve the linearized evolution equations about $(\mu(\cdot,t), \mu^\circ(t)) = S(t)(\mu_0, \mu_0^\circ)$: 
\begin{align}
\p_t (
		\xi_L(\cdot, t),
		\xi_L^\circ(t)
)
	&= [DF(\mu(\cdot,t), \mu^\circ(t))]
	(
		\xi_L(\cdot, t),
		\xi_L^\circ(t)
	)
\end{align}
with initial conditions $(\xi_L(\cdot,0), \xi_L(\cdot,0)) = (\xi_0, \xi_0^\circ)$. In particular, if $(\nu, \nu^\circ) \in L^2(0,1) \times \bbR$ is an equilibrium point of \eqref{eq:evolution}, then 
\begin{align}
	[D\Phi(\nu, \nu^\circ)](\xi_0, \xi_0^\circ) = 
	e^{\cl L}(\xi_0, \xi_0^\circ) \label{eq:dphieq}
\end{align}
where $\cl L = DF(\nu, \nu^\circ)$ is the Fr\'echet derivative of $F$ discussed in Section 3 (see \eqref{eq:linearizedaboutequ}). 

Let $(\mu_0, \mu_0^\circ) \in L^2(0,1) \times \bbR$. By Lemma \ref{l:compact}, \eqref{eq:omequal} and Lemma \ref{l:om} the trajectory $\{ \Phi^n(\mu_0, \mu_0^\circ)\}_{n = 0}^\infty$ is precompact in $L^2(0,1) \times \bbR$ and $\omega_{\Phi}(\mu_0, \mu_0^\circ)$ is a nonempty subset of equilibrium points \eqref{eq:evolution} i.e. fixed points of $\Phi$. Let $(\nu, \nu^\circ) \in \om_{\Phi}(\mu_0, \mu_0^\circ)$ be an equilibrium point. Then by Proposition \ref{p:spectrum} and \eqref{eq:dphieq}
\begin{align}
	\sigma(D\Phi(\nu, \nu^\circ)) = e^{\sigma(\cl L)} = 
	\sigma^u \cup \{1\} \cup \sigma^s
\end{align} 
where $\sigma^u$ is a finite subset of $(1,\infty)$ and $\sigma^s$ is a closed subset of $[0,1)$. By Theorem \ref{t:bp97}, we conclude that $\om(\mu_0, \mu_0^\circ) = \om_{\Phi}(\mu_0, \mu_0^\circ) = \{(\nu, \nu^\circ)\}$ as desired.  
\end{proof}

\bibliographystyle{plain}
\bibliography{researchbibmech}
\bigskip

\centerline{\scshape K.R. Rajagopal}
\smallskip
{\footnotesize
	\centerline{Department of Mechanical Engineering, Texas A\&M University}
	
	\centerline{College Station, TX 77843, USA}
	
	\centerline{\email{krajagopal@tamu.edu}}
}

\bigskip

\centerline{\scshape C. Rodriguez}
\smallskip
{\footnotesize
	\centerline{Department of Mathematics, University of North Carolina}
	
	\centerline{Chapel Hill, NC 27599, USA}
	
	\centerline{\email{crodrig@email.unc.edu}}
}

\end{document}